\def\etal{{et al.}}
\def\eros{{\sc eros}}
\def\macho{{\sc macho}}
\def\gman{{\sc gman}}
\def\mps{{\sc mps}}
\def\ogle{{\sc ogle}}
\def\planet{{\sc planet}}
\def\lmc{{\sc lmc}}
\def\smc{{\sc smc}}

\def\cpu{{\sc cpu }}
\def\ccd{{\sc ccd}}
\def\peida{{\sc peida}}

\documentstyle[11pt,paspconf,epsf]{article}

\begin{document}

\title{EROS2 microlensing search towards the Magellanic Clouds}

\author{T. Lasserre ~ (on behalf of the EROS collaboration)}
\affil{CEA/Saclay, DSM/DAPNIA/SPP, F-91191 Gif-sur-Yvette, France}

\begin{abstract}
\eros2 is a second generation microlensing experiment operating since mid-1996 
at the European Southern Observatory (ESO) at La Silla (Chile).
We present the two year analysis from our microlensing search  
towards the Small Magellanic Cloud (\smc), and report on the intensive
observation of the caustic crossing event \macho-\smc98-1 and the limit derived on
the location of the lens. 
We also give preliminary results from our search towards the Large 
Magellanic Cloud (\lmc); 25~square degrees are being analyzed and two candidates
have been found. This allows us to set another limit on the halo mass fraction
comprised of compact objects.
\end{abstract}

\keywords{Galaxies: halos, kinematics and dynamics, stellar content -- 
Cosmology: dark matter, gravitational lensing}

\section{Introduction}      
A few years after Paczy\'nski's proposal (Paczy\'nski~1986), 
the \eros\ collaboration engaged in long term micro\-lensing
observations towards the Magellanic Clouds in order to probe the Galactic
halo. \eros1 and \macho\ experiments set strong limits 
on the maximum contribution of low mass objects to 
the halo of the Milky Way (Alcock~et~al.~1998).
Towards the \lmc, the optical depth has been estimated by 
\macho\ as $\tau_{\macho}^{\lmc} = 2.9^{+1.4}_{-0.9}\times 10^{-7}$, 
from 8 events (Alcock et al. 1997a);
the time scales associated with these events indicate high mass lenses
 ($\sim 0.5 M_{\odot}$) that are not observed visually. 
Based on 2 candidates, \eros1 gave an upper limit on the halo mass fraction
in \macho s (Ansari et al. 1996) that is below that required to 
explain the rotation curve of our galaxy
\footnote{Assuming the 2 candidates are microlensing events, they correspond to
 $\tau_{\eros}^{\lmc} \sim 0.8 \times 10^{-7}$.}. 
It has been suggested that the lenses might be in the bar/disk of the 
\lmc\ itself (Sahu~1994,Wu~1994); but simple dynamical arguments seem 
to rule out this possibility  (Gould 1995). 
Nevertheless, more complicated \lmc\ models allow for a larger optical
depth ($\sim 1. \, 10^{-7}$).
\smc\ microlensing search provides a test of the halo-lens hypothesis; 
in this model both the optical depth and the typical
durations should be similar towards the \lmc\ and the \smc. 
To date, two \smc\ events have been observed;
they are significantly longer than the average for \lmc\ events. 
However, no definite conclusion can be drawn from this without 
more \smc\ events.

\section{SMC two year analysis}
One candidate, \macho-\smc97-1/\eros-\smc97-1 was found in this analysis
(Alcock~\etal\ ~1997b,~Palanque-Delabrouille~\etal\ ~1998).
The result of a microlensing fit leads to an Einstein~radius crossing time 
$\Delta t=129$~days. The $\chi^2$ is 261~for 279~d.o.f., 
taking into account the $5\%$ intrinsic variability of the amplified star 
($P=5.124$ days, see Palanque-Delabrouille~et~al.~1998, Udalski~et~al.~1997).
This single event allows us to constrain the halo composition,
 in particular we exclude that more than 50~\% of the standard dark halo
is made of $0.5 \:{\rm M}_\odot$ objects.

\section{Results from the caustic crossing in event MACHO-SMC98-1}
The \macho\ collaboration sent a first level alert for this event on  May~ $25$,~1998, 
followed by an announcement that a caustic crossing had occurred on June~ $8$. 
A second caustic crossing was predicted around June~$18$.
After a planned technical maintenance, we could only observe in great 
detail the end of the second caustic crossing. 
Using this data alone, we could extract a limit on the caustic crossing time, which 
together with public data from \macho\ enabled us to determine  (at~a~90\%~ 
likelihood) that the deflector is in the \smc\ (Afonso~et~al.~1998). 
This result has been confirmed and improved by other groups, leading to a  
common publication (Afonso~et~al.~1999, and references therein).

\section{Preliminary analysis of the LMC data}
Since August~1996, we have been monitoring 66~one-square-degree fields towards the \lmc. 
Of these, data prior to May 1998 from 25~square-degrees spread over 
43~fields are being analyzed. This represents 450~Gbytes of raw data, and about 
100~days of \cpu to produce the light curves.
\begin{table}[ht]
\begin{center} 
\begin{tabular}{|l||l||c|c|c|c|c|}
\hline
   & {\em Fit}&$u_0$&$\Delta t \,\,(days)$&$c_{\rm \,bl}\;R$&$c_{\rm \,bl}\;B$&$\chi^2/{\rm d.o.f.}$ \\
\hline
 \lmc-1 & {\em Blended}& $0.23$&$41$&$0.76$& 1 &208/145\\
\hline
 \lmc-2 & {\em Combined}& $0.20$&$106$& 1 & 1 &406/150\\
\hline
\end{tabular}	
\caption{Results of microlensing fits to the candidate \eros2-\lmc-1 and \eros2-\lmc-2. $\Delta t$ is the Einstein radius crossing time, $u_{0}$ is the impact parameter, and $c_{\rm \,bl}\;R(B)$ are the blending coefficients in both colors.}
\label{cand1}	
\end{center} 
\end{table}
About 90~images of each field were taken, with exposure times from 3~min in the center to
12~min in the outermost regions; the sampling is one point every 5~days on average. 
We report a preliminary analysis of the light curves of 17.5~million stars using a new set 
of selection criteria to isolate microlensing candidates. 
Starting from the images we built a star catalog using the \peida\ photometry package, and 
then removed the 90\% most stable stars. 
Among stars with the most significant variations, we used the quality of the 
microlensing fit to select the candidates. 
In order to maximize the number of surveyed stars and to study the background 
of  microlensing searches, we did not remove any star based solely on its position
 in the color-magnitude diagram. 
With this strategy, we characterized the Blue Bumper stars that mimic a microlensing signal.
In this way we removed the stars, located in the upper left of the color-magnitude diagram,
 that pass all cuts, but have the following features :
$ A_{R}-1 > 1.3 (A_{B}-1)$, and 
$A_{R,B} < 1.6$, where $ A_{R(B)}$ are the red(blue) observed amplifications. 
Among the 17.5~million light curves, two events passed all the cuts (see~table~\ref{cand1}).
 Event  \eros2-\lmc-1 is a main sequence star blended in red 
(76\%~of~the~visible~flux~was~magnified).
 Event \eros2-\lmc-2 is located just under the red giant clump, and necessitates a deeper 
photometry study to confirm its validity; it is consistent nevertheless with being achromatic.
\begin{figure}[h]
\begin{center} 
\plotfiddle{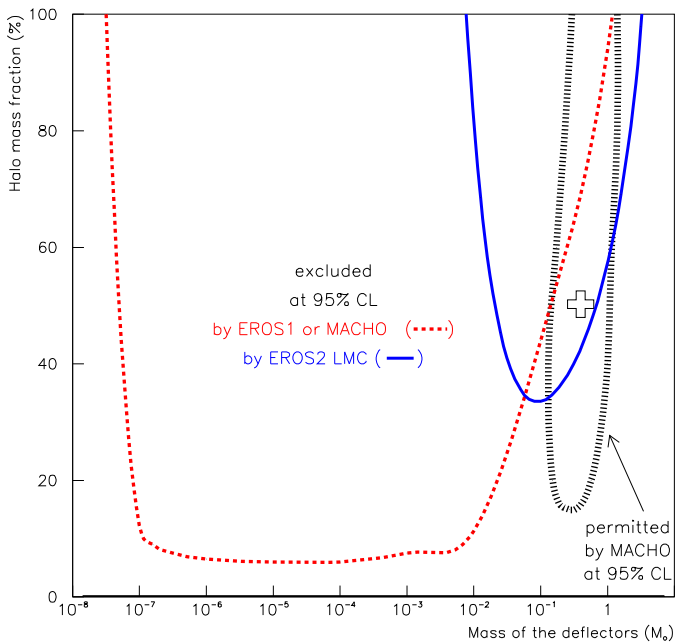}{8. true cm}{0}{50}{50}{-130}{-5}	
\caption{95\% CL exclusion diagram on the halo mass fraction in compact objects
 for the standard halo model. The dashed line indicates the merging of the \macho\
 results (at low mass) and the \eros1 (photographic plates+\ccd) experiments.
 The cross is centered on the area allowed at 95\% CL obtained by the \macho\ \lmc\ 
 two year analysis (Alcock et al. 1997a) (thick dashed line).
 The full line shows the preliminary exclusion limit derived from our \lmc\ analysis.}
 \label{exclusion}
\end{center}
\end{figure}
To set conservative limits on the halo mass fraction $f_{M}$ comprised of compact objects of mass $M$,
we can assume that the observed events are in the dark halo. We only consider 
the standard spherical halo model
described in Palanque-Delabrouille~et~al.~1998.
The most probable mass associated with both candidates is determined by finding the mass
for which the (near~Gaussian) distribution of $log(\Delta t)$ peaks at the geometric mean 
\begin{equation}
\langle \Delta t \rangle = \sqrt{\Delta t_{\lmc-1} \, \Delta t_{\lmc-2}} = 66 \; \mathrm{days}
\label{e:deltat}
\end{equation}
The resulting mass is found to be $1.1 \,{\rm M}_\odot$. We can also define a 68\% confidence 
interval as follows: the upper (lower) bound is determined as the mass for which 16\% 
of detected events would have durations greater (less) than  $\langle \Delta t \rangle$.
This mass interval is found to be :
$M \in [0.15-2.4] \, {\rm M}_\odot$.
Let $N_{M}$ be the total expected number of events for the standard halo model 
(considering~our~detection~efficiency). To be conservative, \
we simply consider our two candidates without taking their mass into account.
In this way, the 95\% CL Poisson limit for a given mass is obtained by computing 
the expected number of events~$C$ compatible with the observations:
$P_{P}(\le 2, C)~=~5\%~\longrightarrow~C~=~6.3$, 
where $P_{P}(a,b)$ is the Poisson probability of observing $a$ events where $b$ are expected. 
The fraction~$f_{M}$ for each mass~$M$ is given by $f_{M}=C/F_{M}$.
This allows us to put a preliminary constraint excluding (at 95~\% CL) 
that 60~\% of the dark halo 
is composed of objects in the range~$[0.02-1] \; \rm{M}_\odot$ (see fig.~\ref{exclusion}).

\section{Discussion}
There is growing evidence, from three different \eros\ data sets 
(\eros1-\lmc, \eros2-\smc, and \eros2-\lmc) 
that the standard spherical halo model fully comprised of $[0.01-2] \; M_{\odot}$ \macho s 
is inadequate. The only way to evade this limit is to suppose that the masses of the \macho s 
are greater than $2 \; M_{\odot}$, or to consider non spherical halos. 
Another way to understand the observed events is to assume that they are due to self-lensing, 
in which case it is important to study their spatial distribution on the face of the \lmc.
In that respect, it is worth noting that our limit is derived from more than 17~million stars spread over 
43~square degrees, in comparison with the \macho\ experiment that monitored 9~millions stars covering 
11~square degrees of the \lmc\ bar (Alcock et al. 1997a).
Finally, more exotic microlensing events (parallax effect, binary lens ...) 
would allow us to locate precisely some lenses, and so 
to test the self-lensing hypothesis.

\end{document}